\documentstyle[preprint,eqsecnum,tighten,prb,aps]{revtex}  
\begin{document}
\draft \preprint{UBCTP-94-003} \title{The Fermi Edge Singularity and
Boundary Condition Changing Operators} \author{ Ian Affleck$^{a}$ and
Andreas W.W. Ludwig$^b$} \address{$^{a}$Canadian Institute for Advanced
Research and Physics Department, University of British Columbia,
Vancouver, B.C.,V6T1Z1, Canada $^b$ Joseph Henry Laboratory, Princeton
University,  Princeton, NJ08544, USA}  \date{\today} \maketitle
\begin{abstract}  The boundary conformal field theory approach to quantum
impurity problems is used to study the Fermi edge singularity, occuring in
the X-ray adsorption probablility.  The deep-hole creation operator, in the
effective low-energy theory, changes the boundary condition on the
conduction electrons.  By a conformal mapping, the dimension of such an
operator is related to the groundstate energy for a finite system with {\it
different} boundary conditions at the two ends. The Fermi edge singularity
is solved using this method,  for the Luttinger liquid including
back-scattering and for the multi-channel Kondo problem.
 \end{abstract} \pacs{ }

\narrowtext \section{Introduction} \label{sec:intro}
	We have recently developed a new method to study quantum impurity
problems, based on conformal field theory with boundaries.\cite{Affleck1}
Using this method we have rederived known results, and in some cases
obtained new results, on the multi-channel\cite{Affleck2} and two-impurity
Kondo problem,\cite{Affleck3} impurities in spin chains,\cite{Eggert}
magnetic monopole-baryon systems\cite{Affleck4} and tunneling in quantum
wires.\cite{Wong}  The purpose of the present paper is to discuss the
extension of our method to another type of quantum impurity problem which
is exemplified by the Fermi edge singularity.\cite{Nozieres}  The standard,
simplified Hamiltonian for this problem is: \begin{equation} H = \sum_{\vec
k}\epsilon_{\vec k}a_{\vec k}^\dagger a_{\vec k} + E_0b^\dagger b +
\sum_{\vec k,\vec k'}V_{\vec k,\vec k'}a_{\vec k}^\dagger a_{\vec k'}b
b^\dagger .\label{Ham} \end{equation} Here $a_{\vec k}$ annihilates a
conduction band electron and $b$ annihilates the ionic deep core electron.
One is interested in calculating the two-point Green's function for the
deep core operator, $b$ and for the operator $ba_{\vec k}^\dagger$ which
creates a  core hole and a conduction electron; the latter transition is
affected by X-ray absorption.  Since the Hamiltonian of Eq. (\ref{Ham})
commutes with $b^\dagger b$, the core electron number, the Hilbert Space
separates into two sectors with the hole absent or present.  When there is
no hole we simply obtain the free conduction electron Hamiltonian:
\begin{equation} H_0 = \sum_{\vec k}\epsilon_{\vec k}a_{\vec k}^\dagger
a_{\vec k}. \label{Ham0}\end{equation}   When the hole is present, the
conduction electrons also feel a scattering potential, $V$:
\begin{equation} H_1 = \sum_{\vec k}\epsilon_{\vec k}a_{\vec k}^\dagger
a_{\vec k}
 + \sum_{\vec k,\vec k'}V_{\vec k,\vec k'}a_{\vec k}^\dagger a_{\vec k'}
.\label{Ham1} \end{equation}
  To calculate the deep core Green's function, $<b^\dagger(t)b(0)>$, we
must solve for the time-dependent response of the conduction electrons to
turning on this potential suddenly at time $0$ and then turning it off
again at time $t$.  As $t\to \infty$ this Green's function exhibits
non-trivial power-law decay, with the exponent depending on the potential,
$V$.

The singularity only depends on the behaviour of the potential right at the
Fermi surface.  Therefore it is entirely determined by the phase shift,
$\delta$, at the Fermi surface, $k_F$.  (In general, it depends on the
phase shifts in all angular momentum channels at $k_F$.  For simplicity,
we  focus on the case where there is only a single channel with a non-zero
phase shift, correponding to a $\delta$-function potential.)  Since the
dependence of the phase shift on $k-k_F$ is irrelevant, the effective
low-energy Hamiltonian for the problem has a constant phase shift for
$-\Lambda <k-k_F <\Lambda$ where $\Lambda$ is the cut-off (which  obeys
$\Lambda <<k_F$).  A constant phase shift corresponds to a simple {\it
boundary condition} relating incoming and outgoing waves at the impurity
location:
\begin{equation}\psi_{\hbox{out}}(0)=e^{2i\delta}\psi_{\hbox{in}}(0).
\end{equation} Thus, the effect of acting with the core hole operator, $b$
is to change the boundary conditions in the low energy theory.  The basic
Fermi edge singularity problem is to calculate the scaling dimension of a
boundary condition changing operator.  As such, we see that it has numerous
generalizations to other quantum impurity problems.

The notion of boundary condition changing operators also plays a
fundamental role in Cardy's  boundary conformal field theory.\cite{Cardy1}
He developed a theory of conformally invariant boundary conditions and
boundary operators.  To each pair of such boundary conditions corresponds a
boundary operator.  However, the reverse is not true.  There are also
boundary operators which {\it do not} change the boundary conditions.  We
have discussed these extensively in our previous papers on quantum impurity
problems.\cite{Affleck1,Affleck2,Affleck3,Eggert,Affleck4,Wong,Affleck5}
The purpose of this paper is to study boundary condition changing operators
in quantum impurity problems using Cardy's boundary conformal field
theory.

The general situation is illustrated in Figure 1.  By s-wave projection, or
its generalizations, we may formulate our quantum impurity problem in one
space and one time dimension, on the half-plane, $r\geq 0$.  Let us assume
that, in the distant past and future some conformally invariant  boundary
conditions A apply.  At (imaginary) time $\tau_1$ a boundary operator,
${\cal O}$, acts which changes the boundary conditions to B.  At
 time $\tau_2$, ${\cal O}^\dagger$ acts and reverts the boundary conditions
to A again.  It is convenient to set the velocity to one and use the
complex coordinate $z\equiv \tau+ir$.  We assume that  the bulk Hamiltonian
is invariant under conformal transformations $z\to f(z)$ and that the
boundary conditions respect the subgroup of conformal transformations that
map the real axis, $z=\tau$ into itself.  It is very useful to apply the
conformal transformation \begin{equation} z=le^{\pi w\over l},
\end{equation} where $l$ is an arbitrary length scale.  This maps the
half-plane into the infinite strip, $w=u+iv$, $0\leq v \leq l$, as shown in
Figure 1. Note that the postive real axis, $\tau >0$ maps onto the lower
boundary of the strip, $v=0$.  For convenience, we have chosen the points
at which the boundary operators act to obey $\tau_i>0$,  so that they both
map onto the lower boundary of the strip.  Hence the boundary conditions on
the top of the strip are always A but on the bottom they are B for $u_1 < u
<u_2$ and A otherwise. It appears that, in all physical cases, the
groundstate with the same boundary condition on the top and bottom of the
strip is the absolute groundstate, the primary state in the conformal tower
corresponding to the identity operator.

We now wish to relate the scaling dimension of the boundary condition
changing operator, $O$, to the groundstate energy of the finite system with
boundary condition A on one side and B on the other.  Letting this
dimension be $x$ and assuming a convenient normalization for the operator,
the Green's function on the half plane with boundary condition $A$ is:
\begin{equation} <A|{\cal O}(\tau_1){\cal O}^\dagger (\tau_2)|A>={1\over
(\tau_1-\tau_2)^{2x}} .\end{equation} The Green's function on the strip is
obtained by the conformal mapping.  Assuming ${\cal O}$ to be primary we
obtain:
 \begin{equation} <AA|{\cal O}(u_1){\cal O}^\dagger
(u_2)|AA>=\left[{(\partial z/\partial w (u_1))(\partial z/\partial w
(u_2))\over [z(u_1)-z(u_2)]^2}\right]^{x}={1\over \left[{2l\over
\pi}\sinh{\pi \over 2l}(u_1-u_2)\right]^{2x}} .\end{equation}  Now let
$u_2-u_1>>l$, giving: \begin{equation} <AA|{\cal O}(u_1){\cal O}^\dagger
(u_2)|AA>\to \left({\pi \over l}\right)^{2x}e^{-\pi
x(u_2-u_1)/l}.\label{GFass}\end{equation}  Here $|AA>$ denotes the
groundstate on the strip with boundary condition $A$ on both sides. It
simply corresponds to the absolute groundstate, as remarked above. On the
other hand, we may also calculate the Green's function on the strip by
inserting a complete set of states: \begin{equation} <AA|{\cal O}(u_1){\cal
O}^\dagger(u_2)|AA>=\sum_n |<AA|{\cal
O}|n>|^2e^{-E_n(u_2-u_1)}.\label{GFexp}\end{equation} Note that, in
general, this sum must include all states with all possible boundary
conditions on the bottom of the strip.  (But a fixed boundary condition $A$
on top of the strip.)  The lowest energy intermediate state is the
groundstate with boundary conditions $A$ on the top and $B$ on the bottom.
Thus \begin{equation} x={l \over
\pi}\left(E_{AB}^0-E_{AA}^0\right).\label{dim-en}\end{equation}  [The other
terms in the expansion of Eq. (\ref{GFexp}) correspond to the contribution
of excited states with boundary conditions $A$ and $B$.]  Thus we see that
the scaling dimension of the operator which changes the boundary conditions
{}from $A$ to $B$ is proportional to the groundstate energy with boundary
conditions $A$, $B$, less the groundstate energy with boundary conditions
$A$, $A$.

So far, we have assumed that ${\cal O}$ produces the groundstate in the
sector of the Hilbert Space with the modified boundary conditions $A$,
$B$.  It can also happen that the lowest energy state produced by ${\cal
O}$ is an excited state in this sector, $E^1_{AB}$.  In that case:
\begin{equation}x={l \over
\pi}\left(E_{AB}^1-E_{AA}^0\right).\label{dim-en2} \end{equation} As we
will see below, the deep core hole operator creates the groundstate with a
modified boundary condition whereas the core hole conduction electron pair
operator (which couples to the X-ray field) produces an excited state with
a modified boundary condition.

In the next section we discuss the Fermi edge singularity for a Fermi
liquid from this perspective, verifying Eqs. (\ref{dim-en}) and
(\ref{dim-en2}).  In Section III we discuss it for a Luttinger liquid; ie.
an interacting one dimensional electron system.  In particular, we confirm
the universal backscattering exponent recently obtained by
Prokof'ev\cite{Prokof'ev} by a different method.
  In Section IV we discuss the multi-channel Kondo/Anderson model from the
perspective of boundary condition changing operators.  In Section V we
discuss the connection between boundary condition changing operators and
fusion, observing that the results of Section III suggest a ``fusion
rules'' approach to the perfectly reflecting fixed point in a Luttinger
liquid.  We hope that this observation may lead to an exact solution of the
mysterious finite reflectance critical points discovered by Kane and
Fisher.\cite{Kane1}

\section{The Fermi edge singularity in a Fermi liquid} In this section we
briefly review the solution of the Fermi edge singularity in a Fermi
liquid, by Schotte and Schotte,\cite{Schotte} and demonstrate that the
relationship between scaling dimensions and energies of Eq.
(\ref{dim-en},\ref{dim-en2}) is obeyed.  For simplicity we assume a
spherically symmetric dispersion relation and only s-wave scattering.  The
problem then reduces to a one-dimensional one defined on the half line
$r>0$.  It is convenient to work with left-moving fermions on the whole
line by reflecting the right-movers to the negative axis: \begin{equation}
\psi_L(-r)\equiv \psi_R(r)\label{LRref}\end{equation}  The left-movers are
functions of $t+x$ only.  (We set the Fermi velocity to one.)  Hence the
Hamiltonian density becomes: \begin{equation} {\cal H}=i\psi^\dagger
{d\over dx}\psi +\delta (x) V\psi^\dagger \psi  bb^\dagger .\end{equation}
We have dropped the ubiquitous subscript $L$. We have also dropped the core
hole energy, $E_0$.  It must be reinserted as a shift in the frequency upon
Fourier transforming the expressions derived below. To make further
progress we bosonize.  We only need consider the left-moving half of a free
massless boson field.  Again we drop the $L$ subscript. The left-moving
fermion is represented in terms of a left-moving boson as: \begin{equation}
\psi \propto e^{i\sqrt{4\pi}\phi}\label{bos}\end{equation}
 The Hamiltonian becomes: \begin{equation} {\cal H}=\left({\partial \phi
\over \partial x}\right)^2-\delta (x){V\over \sqrt{\pi}}{\partial \phi
\over \partial x} bb^\dagger.\end{equation} The two Hamiltonians, $H_0$ and
$H_1$ of Eqs. (\ref{Ham0}) and (\ref{Ham1}) become: \begin{equation} {\cal
H}_0=\left({\partial \phi \over \partial x}\right)^2\end{equation} and
\begin{equation} {\cal H}_1=\left({\partial \phi \over \partial x}-{V\over
2\sqrt{\pi}}\delta (x)\right)^2\end{equation} where we have dropped a
cut-off dependent groundstate energy contribution in ${\cal H}_1$,
corresponding to a shift in the core hole energy, $E_0$.  This might seem
dangerous since the difference in the groundstate energies of $H_0$ and
$H_1$ will play a crucial role in what follows. However, it is only the
{\it universal} part of this groundstate energy, determined by modular
invariance, which will contribute.  Note that ${\cal H}_1$ takes the same
form as ${\cal H}_0$ when written in terms of a shifted field:
\begin{equation} \tilde \phi (x) = \phi (x) -{V\over 4\sqrt{\pi}}\epsilon
(x). \label{phi-tilde}\end{equation}  ($\epsilon (x)=\pm 1$ for $x>0$ and
$x<0$ respectively.)  The fermion field is represented in terms of the
boson as: \begin{equation} \psi (x) \propto e^{i\sqrt{4\pi}\phi
(x)}=e^{i\sqrt{4\pi}\tilde \phi (x)+iV\epsilon
(x)/2}.\label{bostilde}\end{equation}  Recalling, from Eq. (\ref{LRref})
that the field at $x<0$ is the outgoing field, we see that: \begin{equation}
\psi_{\hbox{out}}=e^{2i\delta}\psi_{\hbox{in}},\end{equation} with the
phase shift, \begin{equation} \delta = -V/2\end{equation}  Thus the
bosonized model, with a constant V, produces a $k$-independent phase shift,
equivalent to a boundary condition.

Schotte and Schotte\cite{Schotte} made the crucial observation that ${\cal
H}_1$ is equivalent to ${\cal H}_0$ under a canonical transformation:
\begin{equation} {\cal H}_1=U^\dagger {\cal
H}_0U.\label{unitary}\end{equation}
  This follows from the commutation relations between the left-moving boson
field, $\phi$ and its derivative: \begin{equation} \left[{\partial \phi
(y)\over \partial y},\phi (x)\right]={-i\over 2}\delta
(x-y).\end{equation}  (The validity of this commutator can be checked by
observing that, for a left mover, $\partial \phi /\partial x = \partial
\phi / \partial t$ and recalling that the full boson field is a sum of left
and right parts, both of which contribute to the canonical commutation
relations.)  We find the operator, $U$, satisfying Eq. (\ref{unitary}) is:
\begin{equation} U=e^{2i\delta \phi
(0)/\sqrt{\pi}}.\label{U}\end{equation}  We note that $U$ can only be
considered a unitary operator if we work in the extended Hilbert Space
which includes states with {\it all} possible boundary conditions.  $U$
maps whole sectors of this Hilbert Space, with particular boundary
conditions, into each other.

Now that we understand the boundary condition changing operator, $U$, it is
straightforward to calculate Green's functions.  To calculate the Green's
function for the deep core operator, $b(t)$, we use: \begin{equation}
b^\dagger(t) = e^{iHt}b^\dagger e^{-iHt}=e^{iH_0t}b^\dagger e^{-iH_1t}.
\end{equation} The second equality holds since the core hole must be
present before $b^\dagger$ acts but not after.  The calculation then
reduces to one in the bosonic theory with the core hole operators
eliminated: \begin{equation} <b^\dagger
(t)b(0)>=<e^{iH_0t}e^{-iH_1t}>.\end{equation} Now using, \begin{equation}
e^{-iH_1t}=U^\dagger e^{-iH_0t}U,\end{equation}    we obtain:
\begin{equation} <b^\dagger (t)b(0)>=<U^\dagger (t)U(0)>\end{equation}
Using the free boson propogator: \begin{equation} <\phi (t) \phi (0)> =
-(1/2)\ln (tD),\end{equation} where $D$ is an ultraviolet cut-off, we
obtain: \begin{equation}
 <b^\dagger (t)b(0)>\propto {1\over t^{\delta
^2/\pi^2}}.\label{dimd}\end{equation} Similarly, $b^\dagger (t)\psi (t,0)$
reduces to: \begin{equation} U^\dagger (t)\psi (t,0) \propto
e^{i\sqrt{4\pi}(1-\delta /\pi )\phi (0,t)}, \end{equation} giving the X-ray
edge exponent, \begin{equation}  <b^\dagger (t)\psi (t,0)\psi ^\dagger
(0,0)b(0)>\propto {1\over t^{(1-\delta /\pi
)^2}}\label{dimdpsi}\end{equation}      Fourier transforming, we obtain the
singularity in the X-ray adsorption probability: \begin{equation} \int dt
e^{i(E+E_0)t}<b^\dagger (t)\psi (t,0)\psi ^\dagger (0,0)b(0)>\propto
{1\over (E+E_0)^{\alpha}},\end{equation} with \begin{equation} \alpha =
1-(1-\delta /\pi)^2=2(\delta /\pi)-(\delta /\pi)^2. \end{equation}  (As
mentioned above, strictly speaking, the core hole energy, $E_0$, appearing
on the right hand side should be a renormalized one.)

We now wish to verify Eqs. (\ref{dim-en}) and (\ref{dim-en2}) relating
scaling dimensions to finite-size energy levels.  Thus we put the system in
a box of length $l$ and impose the convenient boundary conditions:
\begin{eqnarray} \psi_R(0)&=&\psi_L(0)\nonumber \\
\psi_R(l)&=&-\psi_L(l).\label{freebc} \end{eqnarray} The former boundary
condition is the result of the s-wave projection from three dimensions.
The latter could arise, for example, from a vanishing boundary condition
for the three dimensional fermions on the surface of a sphere.  It is
actually more convenient to use the alternative formulation of the theory
where we work with left-movers only on an interval of length $2l$ defined
by Eq. (\ref{LRref}).  (When using this formulation we drop the subscript
$L$, as above.) Then the first of Eqs. (\ref{freebc}) simply expresses the
continuity of $\psi_L(x)$ at the origin, whereas the second becomes:
\begin{equation} \psi (-l)=-\psi (l).\end{equation}
 In the free theory, from the bosonization formula of Eq. (\ref{bos}),
taking into account the non-commutativity of the left-moving fields, $\phi
(x)$ and $\phi (y)$, this implies: \begin{equation} \phi (-l)-\phi
(l)=\sqrt{\pi}n, \ \  n=0, \pm 1, \pm 2, ...\end{equation} The mode
expansion for $\phi (x)$, with this boundary condition takes the form:
\begin{equation} \phi (x,t)=\sqrt{\pi}{(t+x)\over 2l}n + \sum_{m=1}^\infty
{1\over \sqrt{2\pi m}}\left[e^{-i\pi {m\over l}(t+x)}a_m +
\hbox{h.c.}\right].\label{mode-exp} \end{equation} Here the $a_m$'s are
boson annihilation operators.  Substituting into the formula for the
Hamiltonian with no core hole, Eq. (\ref{Ham0}), we find the spectrum:
\begin{equation} E=\int_{-l}^l\left({\partial \phi \over \partial
x}\right)^2={\pi \over l}\left[-{1\over 24} +{1\over
2}n^2+\sum_{m=1}^\infty m n_m\right],\label{spectrum}\end{equation} where
the $n_m$'s are the occupation numbers for the $m^{\hbox{th}}$ boson mode,
$a_m^\dagger a_m$. The universal groundstate energy, $E_0=-\pi /24l$, is
required by modular invariance.\cite{Cardy2}It will play an important role
in some of what follows.  We can easily extend this to the theory with
phase shift $\delta$. The first of Eq. (\ref{freebc}) is now modified to:
\begin{equation} \psi_R(0)=e^{2i\delta}\psi_L(0).\end{equation}  We may
again switch to a purely left-moving formulation by defining:
\begin{equation} \psi_L(-x)\equiv e^{2i\delta}\psi_R(x)\ \
(x>0),\end{equation} but now the boundary condition at $x=l$ becomes:
\begin{equation} \psi (-l)=-e^{2i\delta}\psi (l).\end{equation}  In
bosonized form this becomes: \begin{equation} \phi (l)-\phi
(-l)=\sqrt{\pi}(n-\delta /\pi ).\end{equation} Note that this is the
boundary condition obeyed by the shifted field, $\tilde \phi$ of Eq.
(\ref{phi-tilde}). $n$ gets shifted to $n-\delta /\pi$ in the mode
expansion of Eq. (\ref{mode-exp}) and the spectrum becomes:
\begin{equation} E={\pi \over l}\left[-{1\over 24}+{1\over
2}\left(n-{\delta \over \pi}\right)^2+\sum_{m=1}^\infty m
n_m\right],\label{spectrum2}\end{equation}  We can now read off the scaling
dimensions from the spectra of Eq. (\ref{spectrum}) and (\ref{spectrum2})
using Eq. (\ref{dim-en}) and (\ref{dim-en2}).  The operator $U$ or $b$
which creates a core hole takes the groundstate into the phase-shifted
groundstate.  Hence its scaling dimension is: \begin{equation} x_b={l\over
\pi}\left(E^0_{\delta}-E^0_0\right)={\delta^2\over
2\pi^2},\label{dimden}\end{equation} where the superscript denotes the
groundstate and the subscript denotes the phase shift.  The operator
$b\psi^\dagger$ maps the groundstate into the state with phase shift
$\delta$ and one conduction electron present, corresponding to $n=1$ in Eq.
(\ref{spectrum2}).  Hence: \begin{equation} x_{b\psi^\dagger}={l\over
\pi}\left(E^1_{\delta}-E^0_0\right)={1\over 2}\left(1-{\delta \over \pi}
\right)^2.\label{dimdpsien}\end{equation}  Since the the two-point Green's
function for an operator of dimension $x$ scales as $t^{-2x}$ we see that
Eqs. (\ref{dimden}) and (\ref{dimdpsien}) agree with Eqs. (\ref{dimd}) and
(\ref{dimdpsi}).

Note that it was crucial to obtain the correct value of the relative
groundstate energy with different boundary conditions in ordinary to get
the right scaling dimensions.  These groundstate energies emerge naturally
{}from the bosonized form of the theory, as derived above, but are not
obvious from the fermion form.  They can also be derived from a modular
transformation, as discussed in Ref. (\onlinecite{Wong}).

\section{Fermi Edge Singularity in a Luttinger Liquid} In this section we
derive the Fermi edge exponents for a Luttinger liquid, including
backscattering.  Our results, valid for arbitrary strength of the bulk
interactions, agree completely with those of Prokov'ev\cite{Prokof'ev}
derived by a different method.  Our approach is more closely related to
that of Ref. [\onlinecite{Kane2}].  We obtain the universal backscattering
scaling dimension of $1/16$ from the dimension of the ``twist operator'' in
a free boson theory which is related to the order parameter in the
two-dimensional Ising model.  In subsection (A) we discuss the case with
zero backscattering.  In subsection (B) we include backscattering.  In
subsection (C) we consider the case of electrons with spin and general
forward and backward scattering.

The one-dimensional fermions consist of left and right-movers, $\psi_L$ and
$\psi_R$, with general bulk Luttinger liquid interactions.  The (parity
invariant) core hole potential consists of forward and backward scattering
parts, $V_f$ and $V_b$: \begin{equation} {\cal H}_{ch}=\delta (x)
bb^\dagger [V_f(\psi_L^\dagger \psi_L+\psi_R^\dagger
\psi_R)+V_b(\psi_L^\dagger \psi_R+\psi_R^\dagger
\psi_L)]\label{fermH}\end{equation} Once again, it is convenient to
bosonize.  Now we introduce both left and right-moving bosons with:
\begin{eqnarray} \psi_L&\propto &e^{i\sqrt{4\pi}\phi_L}\nonumber \\
\psi_R&\propto &e^{-i\sqrt{4\pi}\phi_R}\nonumber \\ \phi &\equiv& \phi_L +
\phi_R.\label{bosform}\end{eqnarray} The core-hole interaction becomes:
\begin{equation} {\cal H}_{ch}=\delta (x) bb^\dagger \left[-{V_f\over
\sqrt{\pi}}{\partial \phi \over \partial x}+\hbox{constant}\cdot V_b\cos
(\sqrt{4\pi}\phi )\right]\end{equation} The Luttinger liquid interactions
just have the effect of renormalizing the Fermi velocity and rescaling the
boson fields.  Following the notation of Ref. (\onlinecite{Wong}), we
introduce the ``compactification radius'' $R$ for the boson field, such
that $R=1/\sqrt{4\pi}$ in the non-interacting case.  The interactions leave
the bulk Hamiltonian in the non-interacting form
 after the rescaling: \begin{equation} \phi \to \phi /\sqrt{4\pi}R.
\end{equation} Repulsive interactions lead to $R>1/\sqrt{4\pi}$.  The
parameter $R$ is related to the parameter $g$ in Ref. (\onlinecite{Kane1})
and the parameter $\phi$ in Ref (\onlinecite{Prokof'ev}) (not to be
confused with the quantum field in the present notation) by:
\begin{equation} g=e^{2\phi}=1/4\pi R^2. \end{equation} After this
rescaling the full Hamiltonian becomes: \begin{equation} {\cal H}_0={1\over
2}\left({\partial \phi \over \partial t}\right)^2 +{1\over
2}\left({\partial \phi \over \partial x}\right)^2+\delta (x)bb^\dagger
\left[-{V_f\over 2\pi R}{\partial \phi \over \partial
x}+\hbox{constant}\cdot V_b\cos {\phi \over R}\right].\end{equation} [We
have again set the velocity to $1$.] We see that forward and backscattering
have much different effects.  The forward scattering term is always
precisely marginal, leaving the theory non-interacting.  It can be treated
by the same methods as in the Fermi liquid case discussed in the previous
section.  On the other hand, the backscattering term is relevant for
$R<1/\sqrt{4\pi}$ corresponding to repulsive interactions (and irrelevant
in the other case).  It introduces interactions into the boson model,
destroying its harmonic form.

It turns out to be convenient to use a somewhat different basis of fields
to treat this problem.  The same technique has been used previously in our
treatment of this and other quantum impurity problems.  The essential
observation is that we may regard $\phi_R(-x)\equiv \phi_L'(x)$ as a
second  left-moving field.  Such a transformation would presumably not be
very useful if there were any bulk interactions in the bosonic formulation
that mixed $\phi_L(x)$ and $\phi_R(x)$, since these would become
non-local.  But, importantly, the bulk part of the Hamiltonian is free and
decouples into left and right terms.  The only interactions occur at $x=0$
and hence remain local.  Thus we are free to make this transformation if we
wish.  The theory is most conveniently studied in terms of parity even and
odd linear combinations of these two left-moving fields: \begin{equation}
\phi_{e,o}\equiv {1\over \sqrt{2}}\left[\phi_L(x)\mp
\phi_R(-x)\right].\end{equation}  [Note from Eq. (\ref{bosform}) that a
parity transformation has the effect: $\phi_L(x) \leftrightarrow
-\phi_R(-x)$, so the labels $e$ and $o$ are appropriate.]  These fields
obey the commutation relations: \begin{eqnarray}
[\phi_e(x),\phi_e(y)]=[\phi_o(x),\phi_o(y)]&=&{-i\over 4}\epsilon
(x-y)\nonumber \\ {[}\phi_e(x),\phi_o(y)]&=&{-i\over 4}
\label{com}\end{eqnarray} The Hamiltonian  reduces to a sum of  commuting
terms involving the left-moving fields, $\phi_{e,o}$ as: \begin{equation}
{\cal H}=\left({\partial \phi_e \over \partial x}\right)^2+\left({\partial
\phi_o \over \partial x}\right)^2+\delta (x) bb^\dagger \left[-{V_f\over
\sqrt{2}\pi R}{\partial \phi_e\over \partial x}+\hbox{constant}\cdot
V_b\cos {\sqrt{2}\phi_o\over R}\right].\label{Hameo}\end{equation} The
usefulness of this peculiar change of variables is now evident; the
Hamiltonian separates into two commuting terms for forward and backward
scattering.   The original fermion fields, at the origin become:
\begin{equation} \psi_{L,R}(0)\propto \exp \left[i2\sqrt{2}\pi R\phi_e(0)
\right] \cdot \exp \left[\pm i{\phi_o(0)\over \sqrt{2}R}\right]
.\label{boseo}\end{equation}
 The initial boundary conditions on the boson fields, $\phi_e$ and $\phi_o$
simply specify that both fields are continuous at the origin.  We will
argue that the forward and backward scattering terms in the Hamiltonian are
equivalent, at low energies, to modified boundary conditions on the fields
$\phi_e$ and $\phi_o$ respectively.  We will again be interested in
calculating the dimensions of boundary condition changing operators.  These
will simply factorize into products of commuting operators acting on
$\phi_e$ and $\phi_o$ respectively.  All scaling dimensions will be sums of
forward and backward scattering parts from $\phi_e$ and $\phi_o$
respectively.  Similarly, the finite size spectrum will separate into a sum
of even and odd excitation energies. For instance, he scaling dimension of
the fermion field is: \begin{equation} x_\psi = \pi R^2+{1\over 16\pi
R^2}.\end{equation} (Note that this gives $x=1/2$ for $R=1/\sqrt{4\pi}$,
the free fermion case.) \subsection{Forward Scattering} Forward scattering
can be treated by the same method reviewed in Section II.  The boundary
conditon changing operator, $U$, of Eq. (\ref{U}) is: \begin{equation}
U=e^{-iV_f\phi_e(0)/\sqrt{2}\pi R}.\end{equation}  Thus, if we only have
forward scattering, the deep hole annihilation operator has scaling
dimension: \begin{equation} x_b={V_f^2\over
16\pi^3R^2}.\label{xd}\end{equation}  Similarly the operator $b^\dagger
\psi$, determining the X-ray exponent becomes: \begin{equation} U^\dagger
\psi_{L,R} \propto \exp \left[i\left({V_f\over \sqrt{2}\pi R} +
2\sqrt{2}\pi R\right)\phi_e(0)\right] \cdot \exp \left[\pm i{\phi_o(0)\over
\sqrt{2}R}\right] ,\end{equation} of dimension: \begin{equation} x = \pi
\left( R+{V_f\over 4\pi^2 R}\right)^2+{1\over 16\pi R^2}.\label{xdpsi}
\end{equation}

Again this can be obtained from the finite size spectrum with phase-shifted
boundary condition.  We begin by defining the system on a circle of
circumference $2l$;  $-l<x<l$.  Antiperiodic boundary conditions are
imposed on the fermion fields: $\psi_{L,R}(l)=-\psi_{L,R}(-l)$.  From the
bosonization formula, Eq. (\ref{bosform}), taking into account the
commutation relations, it can be seen\cite{Wong} that the corresponding
boundary conditions on the fields, $\phi_{e,o}$ are: \begin{eqnarray}
\phi_e(l)-\phi_e(-l)&=&\pi \sqrt{2}Rn \nonumber \\
\phi_o(l)-\phi_o(-l)&=&{m\over 2\sqrt{2}R},\label{eobc} \end{eqnarray}
where $n$ and $m$ are arbitrary integers obeying: \begin{equation} n+m=0 \
\ \ (\hbox{mod}\ 2).\end{equation}

The forward scattering interaction of Eq. (\ref{Hameo}) shifts the boundary
condition on $\phi_e$ to: \begin{equation} \phi_e(l)-\phi_e(-l)=\pi
\sqrt{2}Rn+{V_f\over 2\sqrt{2}\pi R},\end{equation} as can be seen by
absorbing the potential scattering into a redefinition of the field
$\phi_e$ as in Eq. (\ref{phi-tilde}).  Equivalently, the modified boundary
conditions on the fermion fields are: \begin{eqnarray} \psi_L(0^+)&=&\exp
\left[iV_f/4\pi R^2 \right]\psi_L(0^-)\nonumber \\ \psi_R(0^-)&=&\exp
\left[iV_f/4\pi R^2 \right]\psi_R(0^+).\end{eqnarray}  The mode expansions
for the left-moving fields, $\phi_{e,o}$ are: \begin{eqnarray}
\phi_e(t+x)&=&{(t+x)\over 2l}\left(\sqrt{2}\pi Rn+{V_f\over 2\sqrt{2}\pi
R}\right)+...\nonumber \\ \phi_o(t+x)&=&{(t+x)\over 2l}{m\over
2\sqrt{2}R}+... \ \ \ \  [n+m=0 \ \ \ (\hbox{mod}\ 2)],\end{eqnarray} where
the $...$ represents the harmonic modes.  The finite-size spectrum is given
by: \begin{equation} E={\pi \over l}\left[-{1\over 12}+ \pi
\left(Rn+{V_f\over 4\pi^2R}\right)^2+{m^2\over 16\pi R^2}+ ... \right]\ \
\  [n+m=0 \ \ \ (\hbox{mod}\  2)].\label{spec}\end{equation}  Note that the
universal groundstate energy is doubled since there are two boson fields,
$\phi_e$ and $\phi_o$.

To make contact with the general formalism of Sec. I, we ``fold'' the
system about $x=0$, regarding the left-moving fields at $x<0$ as
right-moving fields at $x>0$.  The above spectrum then corresponds to
having a trivial boundary condition at $x=l$ and a phase shift at $x=0$
corresponding to the forward scattering potential, $V_f$.  The groundstate
energy ($n=m=0$), $El/\pi\equiv x=V_f^2/16\pi R^2$, gives the dimension of
the deep hole creation operator, $b$, in agreement with Eq. (\ref{xd}).
The excited state with $n=m=1$ gives the dimension of the operator
$b^\dagger \psi$, \begin{equation} x_{b^\dagger \psi}=\pi \left(R+{V_f\over
4\pi^2R}\right)^2+{1\over 16\pi R^2},\end{equation} as in Eq.
(\ref{xdpsi}). Note that the energies are sums of even and odd parts and
only the even part is modified by forward scattering. \subsection{Back
Scattering}
  Rather different methods are required to treat the back scattering
interaction in Eq. (\ref{Hameo}).  The Hamiltonian is presumably not
unitarily equivalent to the free one.  Furthermore, the backscattering
interaction is relevant, for $R<1/\sqrt{4\pi}$, and is expected to
renormalize to large values at low energies.  Hence the scaling dimensions
that we are after will not depend on the actual value of $V_b$.  A way
around these difficulties was found by Prokof'ev.\cite{Prokof'ev} He argued
that it should be possible to replace the cosine back-scattering
interaction by one quadratic in $\phi_o$ and exhibited a unitary operator,
independent of $V_b$, which, at low energies, reduced the Hamiltonian to
the non-interacting one.  An alternative approach, which we use here, is to
focus on the finite-size spectrum.  As argued by Kane and
Fisher,\cite{Kane1}[(see also Ref. (\onlinecite{Eggert})] $V_b$
renormalizes to $\infty$ corresponding to perfectly reflecting behaviour at
low energies.  The problem then is to find the scaling dimension of the
operator which changes a perfectly transmitting boundary condition
($V_b=0$) into a perfectly reflecting one ($V_b=\infty$).  (For the moment,
we set the forward scattering to zero.)  As explained in Sec. I, this is
equivalent to finding the groundstate energy with a perfectly transmitting
boundary at one end of the system and a perfectly reflecting one at the
other.  This problem, which is essentially trivial, is solved in Ref.
(\onlinecite{Wong}), Appendix B. [See also Ref. (\onlinecite{Affleck4}).]
It is simplest to go back to the original left and right moving bosons of
Eq. (\ref{bosform}) on the interval $-l\leq x \leq l$ with $-l$ and $l$
identified. The perfectly reflecting boundary condition at the origin is:
\begin{equation} \phi_L(0^\pm )=-\phi_R(0^\pm).\end{equation} Note that the
spatial component of the current is $J_1\propto \partial \phi /\partial t$
and vanishes at $x\to 0$ as the origin is approached from either side.  We
may now use the trick of regarding the right-movers as reflected
left-movers on {\it both} the postive and negative side of the origin.
Hence the system becomes equivalent to a single left-mover on an interval
of length $4l$, with boundary conditions: \begin{equation} \phi (4l)-\phi
(0) = 2\pi nR.\end{equation}
 The spectrum is: \begin{equation}
 E={\pi \over 2l}\left[-{1\over 24}+ 2\pi R^2n^2+\sum_{m=1}^\infty mn_m
\right],\label{specbs}\end{equation} Note the extra prefactor of $1/2$ due
to the doubled length and the single groundstate energy term $-1/24$ since
we effectively only have one boson field. From this formula and Eq.
(\ref{spec}), we can read off the scaling dimension of the operator which
converts the transmitting boundary condition into a reflecting one:
\begin{equation} x_b=-1/48+1/12=1/16.\end{equation}  Alternatively, in
terms of the even and odd bosons, the perfectly reflecting boundary
condition is: \begin{eqnarray} \phi_e(0^+)&=&\phi_e(0^-)\nonumber \\
\phi_o(0^+)&=&-\phi_o(0^-).\end{eqnarray} $\phi_e$ obeys a trivial boundary
condition but $\phi_o$ obeys a twisted one.  We see that the  operator
which changes the boundary condition from transmitting to reflecting acts
trivially on $\phi_e$ but changes the sign of $\phi_o$. We see that
inserting this boundary condition changing operator at the points $\tau_1$
and $\tau_2$, corresponds to inserting a cut along the $\tau$ axis between
these two points.  The field $\phi_o$ changes sign accross the cut.  This
boundary condition changing operator is known as a {\it twist operator}.
Its dimension, $1/16$, is calculated in a somewhat different way in Ref.
(\onlinecite{Ginsparg}) where its relationship with the Ising model order
parameter is also discussed.  We note that the term in Eq. (\ref{specbs})
proportional to $n^2$ is identical to a term in the spectrum with no
scattering, Eq. (\ref{spec}) with $V_f=0$.  This, together with the $m$
even terms in Eq. (\ref{specbs}), can be identified as the contribution to
the energy from $\phi_e$, which is unchanged by back scattering.  On the
other hand, the contribution of $\phi_o$ is changed significantly; in
particular, it becomes independent of $R$.

The operator $b^\dagger \psi$ must correspond to a state with $n=1$, since
the $\phi_e$-dependence of this operator is simply that of the free
fermion.  Hence: \begin{equation} x_{b^\dagger \psi}=1/16 + \pi
R^2.\end{equation}  We may now include both forward and back scattering.  A
minor extension of the above calculation shows that the integer $n$,
referring to the even sector is shifted exactly as before, giving:
\begin{equation}
 E={\pi \over 2l}\left[-{1\over 24}+ 2\pi R^2\left(n+{V_f\over
4\pi^2R^2}\right)^2+\sum_{m=1}^\infty mn_m
\right].\label{specbfs}\end{equation}  Thus with both forward and back
scattering, the dimensions of $d$ and $b^\dagger \psi$ become:
\begin{eqnarray} x_b&=&{1\over 16} + {V_f^2\over 16\pi^3R^2}\nonumber \\
x_{b^\dagger \psi}&=&{1\over 16} + \pi \left(R+{V_f\over
4\pi^2R}\right)^2.\end{eqnarray} These results agree completely with those
of Prokof'ev [ Eq. (24) and (30) of Ref. (\onlinecite{Prokof'ev}].

\subsection{Including Spin} We now extend the model to include the electron
spin.  The core hole part of the Hamiltonian is still given by Eq.
(\ref{fermH}) but with an implicit sum over spin indices.  In general, we
allow two Luttinger liquid interactions which preserve only the $U(1)$
subgroup of the spin rotation group corresponding to rotations about the
$z$-axis.  We again bosonize and change variables from independent bosons
for spin up and down to spin and charge bosons, $\phi_s$ and $\phi_c$.
[See Ref. (\onlinecite{Wong}) for further details.]  We again introduce
parity even and odd left-moving fields, for both spin and charge.  The full
Hamiltonian density, in bosonized form, becomes: \begin{eqnarray} {\cal
H}&=&\left({\partial \phi_{c,e} \over \partial x}\right)^2+\left({\partial
\phi_{c,o} \over \partial x}\right)^2 +\left({\partial \phi_{s,e} \over
\partial x}\right)^2+\left({\partial  \phi_{s,o} \over \partial
x}\right)^2\nonumber \\ &&+\delta (x) bb^\dagger \left[-{\sqrt{2}V_f\over
\pi R_c}{\partial \phi_{c,e}\over \partial x}+\hbox{constant}\cdot V_b\cos
{\sqrt{2}\phi_{c,o}\over R_c}\cos {\sqrt{2}\phi_{s,o}\over
R_s}\right]\end{eqnarray} Here the ``compactification radii'' for charge
and spin bosons are determined by the Luttinger liquid interactions.  They
are related to the parameters of Kane and Fisher by: \begin{equation}
g_{\rho ,s} = 1/\pi R^2_{c,s}.\end{equation}   Prokof'ev only considers the
$SU(2)$ invariant case, $R_s=1/\sqrt{2\pi}$ and parameterizes the charge
sector interactions by: \begin{equation} e^{2\phi}=1/2\pi
R_c^2.\end{equation}
 Note that the forward scattering term only involves the even charge boson,
$\phi_{c,e}$ and is, again, marginal.  The back scattering term involves
the odd charge and spin bosons.  Under renormalization we expect to
generate pure spin and charge terms, $\cos [ 2\sqrt{2}\phi_{o,s}/R_s]$ and
$\cos [2\sqrt{2}\phi_{o,c}/R_c]$.  The phase diagram is discussed
extensively by Kane and Fisher.\cite{Kane1} [See also Ref.
(\onlinecite{Wong}).]  Depending on the bulk interaction parameters, $R_c$
and $R_s$, there are four stable phases, in which spin and charge are
either perfectly reflected or perfectly transmitted.  For some values of
$R_c$ and $R_s$ more than one of these phases is stable for some range of
scattering parameters and some unstable non-trivial fixed points occur at
intermediate scattering potential.  We restrict our attention to the four
stable phases.  The fermion operators at the origin can be written:
\begin{eqnarray} \psi_{\uparrow ,L}&\propto & \exp \left\{
i\left[\sqrt{2}\pi R_c\phi_{c,e}+\sqrt{2}\pi R_s\phi_{s,e}+{\phi_{c,o}\over
\sqrt{2}R_c}+{\phi_{s,o}\over \sqrt{2}R_s}\right]\right\}\nonumber \\
\psi_{\uparrow ,R}&\propto & \exp \left\{ i\left[\sqrt{2}\pi
R_c\phi_{c,e}+\sqrt{2}\pi R_s\phi_{s,e}-{\phi_{c,o}\over
\sqrt{2}R_c}-{\phi_{s,o}\over \sqrt{2}R_s}\right]\right\}\nonumber \\
\psi_{\downarrow ,L}&\propto & \exp \left\{ i\left[\sqrt{2}\pi
R_c\phi_{c,e}-\sqrt{2}\pi R_s\phi_{s,e}+{\phi_{c,o}\over
\sqrt{2}R_c}-{\phi_{s,o}\over \sqrt{2}R_s}\right]\right\}\nonumber \\
\psi_{\downarrow ,L}&\propto & \exp \left\{ i\left[\sqrt{2}\pi
R_c\phi_{c,e}-\sqrt{2}\pi R_s\phi_{s,e}-{\phi_{c,o}\over
\sqrt{2}R_c}+{\phi_{s,o}\over \sqrt{2}R_s}\right]\right\}.\end{eqnarray}

We first consider the case of forward scattering only.  The unitary
operator which eliminates the forward scattering term from the Hamiltonian
is: \begin{equation} U = e^{-i\sqrt{2}V_f\phi_{c,e}(0)/\pi
R_c}.\end{equation} Hence, with no back scattering: \begin{eqnarray}
x_b&=&{V_f^2\over 4\pi^3R_c^2} \nonumber \\ x_{b^\dagger \psi}&=&{1\over
4\pi}\left[\left({V_f\over \pi R_c}+\pi R_c\right)^2+\left({1\over
2R_c}\right)^2+(\pi R_s)^2+\left({1\over
2R_s}\right)^2\right].\end{eqnarray} The back scattering term, depending on
the values of $R_c$ and $R_s$, can twist the charge or spin odd boson; ie.
produce a perfectly reflecting boundary for charge and/or spin.  From the
discussion of the previous section, and from the finite-size spectra in
Ref. (\onlinecite{Wong}), we see that each twist operator has dimension
$1/16$.  Hence, $x_b$ is increased by $1/16$ in the charge-transmitting,
spin-reflecting or charge-reflecting, spin-transmitting cases and by $1/8$
in the charge and spin reflecting case.  Only the last case was considered
by Prokof'ev corresponding to $SU(2)$ symmetry, $R_s=1/\sqrt{2\pi}$ and a
repulsive Luttinger liquid interaction, $R_c>1/\sqrt{2\pi}$.  Similarly to
what happened in the spinless case, and as seen from the finite size
 spectrum in Ref. (\onlinecite{Wong}), the dimension of $x_{b^\dagger
\psi}$ is modified for relevant backscattering by the replacement of the
term $1/16\pi R^2$ by $1/16$.  This replacement is made for the $R_c$ term
in the case where charge is perfectly reflected and is made for the $R_s$
term in the case where spin is perfectly reflected.  In particular, when
charge and spin are both perfectly reflected: \begin{equation}
x_{b^\dagger \psi}={1\over 4\pi}\left[\left({V_f\over \pi R_c}+\pi
R_c\right)^2+(\pi R_s)^2\right]+{1\over 8}.\end{equation}  This is, again,
in perfect agreement with Prokof'ev.\cite{Prokof'ev} \section{The Kondo
Effect} In the previous sections we have considered fairly trivial examples
of boundary condition changing operators where the Hamiltonian, $H_1$ which
occurs after a deep hole is created, differs from the Hamiltonian, $H_0$
which occurs without a deep hole, by a simple potential scattering term.
More general possibilities also exist in which the Hamiltonian after the
creation of a deep hole differs from the unperturbed Hamiltonian by some
non-trivial interaction terms involving additional dynamical degrees of
freedom at the deep hole location.  An instructive example of this is
provided by the Kondo/Anderson model.  ie. we consider the Kondo
Hamiltonian but represent the impurity spin in terms of a deep hole fermion
operator with spin, $b_\alpha$: \begin{equation} \vec
S_{\hbox{imp}}=b^{\alpha \dagger}{\vec \sigma^\beta_\alpha \over 2}b_\beta
.\end{equation}
  The Hamiltonian is thus: \begin{equation}H = \sum_{\vec k}\epsilon_{\vec
k}a_{i \vec k}^\dagger a_{i \vec k} +E_0b^{\alpha \dagger}b_\alpha
+Jb^{\alpha \dagger}{\vec \sigma^\beta_\alpha \over 2}b_\beta \cdot
\sum_{\vec k,\vec k'}a_{i \vec k}^{\gamma \dagger} {\vec
\sigma^\delta_\gamma \over 2} a_{i \delta \vec k'}.\end{equation} Here $i$
is a channel index which runs over $k$ values, corresponding to the
multi-channel Kondo effect. We may again consider the X-ray edge problem
for this core hole state which now carries spin.  Again the Hamiltonian
commutes with the core hole occupation number.  When the core state is
either vacant or doubly occupied the Hamiltonian reduces to the free one.
When it is singly occupied the Hamiltonian reduces to the Kondo model.  We
have discussed the Kondo model extensively, emphasizing that at low
energies the impurity spin is screened and all that remains is an effective
conformally invariant boundary condition on the low energy electronic
degrees of freedom.\cite{Affleck5,Affleck2}  Thus, at low energies,
$b_\alpha$ is again a boundary condition changing operator. However,
 in the overscreened case ($k>2s_{\hbox{imp}}$) this Kondo boundary
condition is of a non-trivial type leading to exotic effects like
fractional groundstate degeneracy and non-trivial scaling laws.  In all our
previous discussion of the Kondo effect, we only considered correlation
functions involving the conduction electron operators and the impurity spin
operator, $\vec S_{\hbox{imp}}$; not the core hole operator $b_\alpha$.
All the corresponding boundary operators live in the Hilbert Space with a
fixed, Kondo, boundary condition.  Equivalently, they correspond to states
in the finite size spectrum with Kondo boundary conditions at both ends of
the system. In this section we enlarge our discussion to include the
boundary condition changing operator $b_\alpha$.

The boundary operator corresponding to $b_\alpha$ can be identified from
the finite size spectrum with Kondo boundary conditions at one end and free
boundary conditions at the other.  The groundstate with free boundary
conditions at both ends is the spin and flavour singlet charge zero state.
The Kondo boundary condition is obtained by fusion with the spin $1/2$
flavour singlet charge zero operator and therefore the resulting spectrum
contains the corresponding conformal tower.\cite{Affleck5}  Clearly the
corresponding primary field, $g_\alpha$,  has the right quantum numbers to
correspond to the deep hole operator, $b_\alpha$.  Hence we conclude
that\cite{Affleck5} \begin{equation} x_b={3/4\over 2+k},\end{equation} the
dimension of the Kac-Moody primary field, $g_\alpha$. This result was
suggested previously by Tsvelik in another context.\cite{Tsvelik} In the
special case of one channel, $x_b=1/4$.  In this case the Kondo fixed point
is of Fermi liquid type and simply corresponds to a $\pm \pi /2$ phase
shift for spin up or spin down electrons.  This corresponds exactly with
the results of Sec. II.  From Eq. (\ref{dimden}) the scaling dimension is:
\begin{equation} x_b={1\over 2}\left({\delta_\uparrow \over \pi}\right)^2
+{1\over 2}\left({\delta_\downarrow \over \pi}\right)^2={1\over
2}\left({1\over 2}\right)^2+{1\over 2}\left({1\over 2}\right)^2={1\over
4}.\end{equation}

We may also calculate the dimension of the ``X-ray operators'', $b^{\alpha
\dagger}\psi_{\beta i}$.  These have charge $Q=1$, transforms under the
fundamental representation of the flavour group and have spin $j=0$ or $1$
depending on how the spin indices are contracted.  A primary field of these
quantum numbers is obtained by fusion of the $j=1/2$ primary with the
conformal tower of the free electron operator, with $j=1/2$, $Q=1$ and
fundamental flavour representation, for $k>2$.  We expect it to correspond
to the ``X-ray operators'' which therefore have dimension:\cite{Affleck5}
\begin{equation} x_j = {1\over 4k} + {k^2-1 \over2k( 2+k)}+ {j(j+1)\over
2+k},\end{equation} with $j=0$ or $1$.  In the special case, $k=1$,
$x_0=1/4$.  In this case the $j=1$ primary doesn't exist so $x_1$ must be a
descendent with $x_1=5/4$.  These results can be obtained from the phase
shift picture of Eq. (\ref{U}), using: \begin{eqnarray} b_{\uparrow
\downarrow} &\propto&  e^{2i(\delta_\uparrow
\phi_\uparrow+\delta_\downarrow \phi_\downarrow )/\sqrt{\pi}} \propto
e^{\pm i\sqrt{2\pi}\phi_s} \nonumber \\ \psi_{\uparrow \downarrow}
&\propto& e^{ i\sqrt{2\pi}\phi_c}e^{\pm i\sqrt{2\pi}\phi_s},\end{eqnarray}
where $\phi_c$ and $\phi_s$ are charge and spin bosons and we have chosen
$(\delta_\uparrow , \delta_\downarrow )=(\pi /2,-\pi /2)$ in $b_{\uparrow}$
and  $(\delta_\uparrow , \delta_\downarrow )=(-\pi /2,\pi /2)$ in
$b_{\downarrow}$. \section{Fusion}  A useful method of generating new
conformally invariant boundary conditions from known ones is by
fusion.\cite{Cardy1}  We used this method to identify the multi-channel
Kondo boundary conditions.\cite{Affleck5}  On the other hand, we did not
use this technique to find the boundary conditions corresponding to perfect
reflection in a Luttinger liquid.\cite{Wong}  The insights gained from this
paper suggest that such an approach is, in fact, possible.  We hope that
this will prove useful in solving a major open problem: the finite
reflectance critical points of Kane and Fisher.\cite{Kane1}

If boundary condition (b.c.) $B$ is obtained from b.c. $A$ by fusion with
some operator ${\cal O}^a$ then the partition function on a finite cylinder
with b.c. $B$ at one end and an arbitrary b.c. $C$ at the other, $Z_{BC}$,
is determined by the same partition function with $B$ replaced by $A$,
$Z_{AC}$,  together with the fusion rule multiplicities $N^{ab}_c$.
[$N^{ab}_c$ is a non-negative integer which specifies how many times the
primary field ${\cal O}^c$ appears in the operator product expansion of
${\cal O}^a$ with ${\cal O}^b$.]  The partition functions can be expanded
in characters, $\chi_a$ of the $a^{\hbox{th}}$ conformal tower,
\begin{equation} Z_{AB}=\sum_an^a_{AB}\chi_a ,\end{equation} where the
$n^a_{AB}$'s are non-negative integers.  $Z_{BC}$ is determined by:
\begin{equation} n_{BC}^b=\sum_cN^{ab}_cn^c_{AC}.\end{equation}

As first shown by Cardy\cite{Cardy1}, under certain circumstances if an
operator ${\cal O}^a$ changes b.c. $A$ into b.c. $B$ (ie. produces the
groundstate on the strip with b.c. $B$) then $B$ can be obtained from $A$
by fusion with the operator  ${\cal O}^a$.  We saw an example of this in
the previous section where the $j=1/2$ primary operator changed the free
b.c. into the ``Kondo b.c.'' and the latter was
   obtained from the former by fusion with this operator. Another example
is provided by our discussion of forward scattering in  Sec. III.  With no
scattering general primary boundary operators are of the form:
\begin{equation}  \exp \left[in2\sqrt{2}\pi R\phi_e(0) \right] \cdot \exp
\left[im{\phi_o(0)\over \sqrt{2}R}\right].\end{equation} After fusion with
\begin{equation} U^\dagger=e^{iV_f\phi_e(0)/\sqrt{2}\pi R},\end{equation}
we obtain the boundary operators: \begin{equation}  \exp
\left[i(n2\sqrt{2}\pi R+V_f/\sqrt{2}\pi R)\phi_e(0) \right] \cdot \exp
\left[im{\phi_o(0)\over \sqrt{2}R}\right],\end{equation} corresponding to
the finite size spectrum of Eq. (\ref{spec}).

 We discovered in Sec. III that the operator which changes a perflectly
transmitting boundary condition into a perfectly reflecting one is the
twist operator of dimension $x=1/16$ for the odd boson, $\phi_o$.
Presumably the latter boundary condition can be obtained from the perfectly
transmitting one by fusion with this twist operator.  (We have checked this
explicitly for certain  values of the compactification radius, $R$.)
Including spin we have two different twist operators for $\phi_{c,o}$ and
$\phi_{s,o}$. Presumably  fusion with some generalization of these twist
operators will give the finite reflectance critical point.

IA would like to thank Junwu Gan, Gabriel Kotliar and Chandra Varma for
interesting him in this problem and Eugene Wong for helpful discussions.
AWWL thanks Thierry Giamarchi for discussions  at the early stages of this
 work. This research is supported in part by NSERC of Canada.

 \begin{figure} \caption{Boundary condition changing
operators act at times $\tau_1$ and $\tau_2$.}\end{figure} \end{document}